\documentclass[12pt]{article}
\usepackage{pic04}
\usepackage{hyperref}
\usepackage{url}
\usepackage{graphicx}

\begin{document}

\title{\bf X-RAY SCANNER FOR ATLAS BARREL TRT MODULES}
\author{
Taeksu Shin, O. K. Baker, K. W. McFarlane, Vassilios Vassilakopoulos\\
{\em Department of Physics, Hampton University, Hampton VA 23668} \\
}
\maketitle

\baselineskip=14.5pt
\begin{abstract}
X-ray scanners for gain mapping of ATLAS Barrel Transition Radiation Tracker
(TRT) modules \footnote{ Supported in part by the U.S. NSF under awards 
PHY007268 and PHY0114343 (COSM).} were developed at Hampton University for 
quality assurance purposes. Gas gain variations for each straw of the TRT 
modules were used to decide whether wires should be removed or restrung, 
and to evaluate overall module quality.
\end{abstract}

\baselineskip=17pt

\section{Introduction}
In the ATLAS experiment \cite{atlastech} at the Large Hadron Collider (LHC)
at CERN, straw-tube Transition Radiation Tracker (TRT) Barrel detectors 
will play an major role in providing precise tracking with transition 
radiation (TR) capability for electron identification.  To achieve 
stability against high-voltage breakdown at the very high interaction rate
of the LHC, it is important that the sense wires be centered in the
1.44 m-long straws.  An off-center (offset) wire will break down at a lower 
high
voltage than a centered wire, reducing the range of operation of the
device.  Wires can be offset as a function of position for a variety of
reasons, including bending of the 1.44 m-long straws.  Since the average gas
gain also increases with offset, a measurement of gas gain as a function
of position along a wire can be used to indicate whether or not the wire is 
offset at any point \cite{duke}. Here, we briefly describe the common features 
of two X-ray scanners built at Hampton University to determine whether wires 
are offset.

\section{X-ray Scanner for ATLAS TRT}
The X-ray photon energy for the gain survey can be chosen to optimize the
time to obtain the required gas gain resolution. This depends on several 
factors: attenuation of the X-rays in the module, conversion of the X-rays in 
the active gas, and ability to find the gain through an analysis of the 
spectrum. For surveying the largest of the three module types
(type 3), we found that 12 keV is close to optimum. To obtain a photon
beam of this energy, X-ray fluorescence from bromine was used; the primary
beam came from a 50 kV x-ray tube.

\begin{figure}[hbtp]
  \centerline{
   \hbox{ 
     \hspace{0.1cm}
     \includegraphics[width=6.5cm, height=3.5cm]{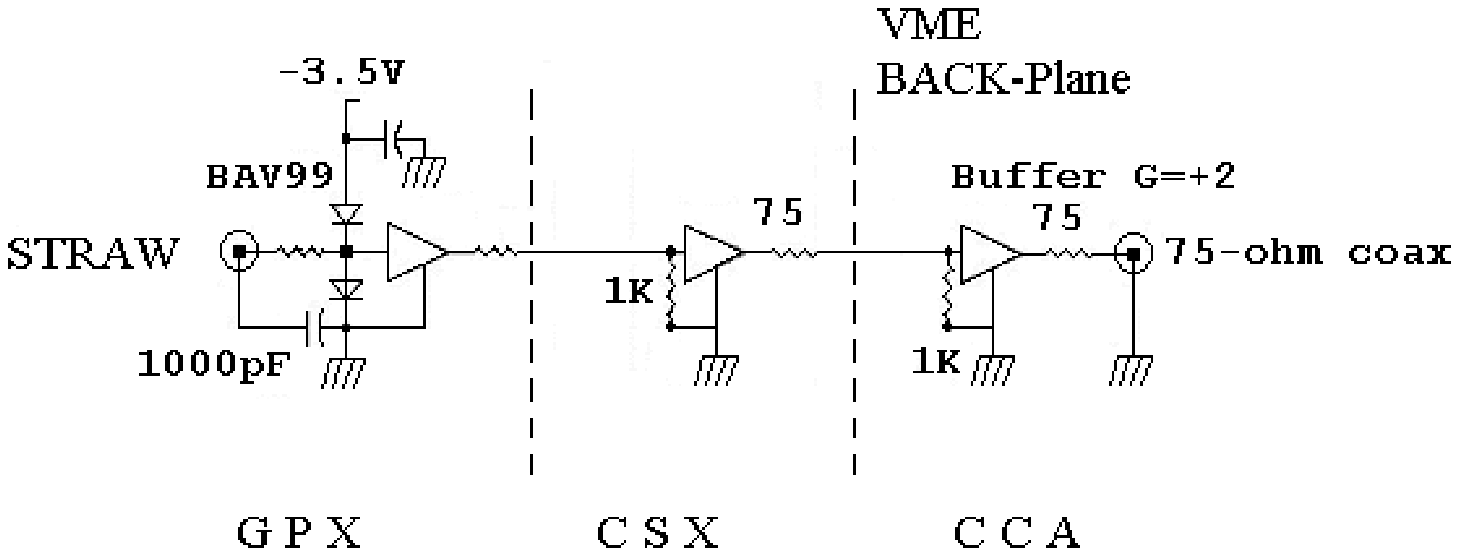}
     \hspace{0.1cm}
     \includegraphics[width=6.5cm, height=3.5cm]{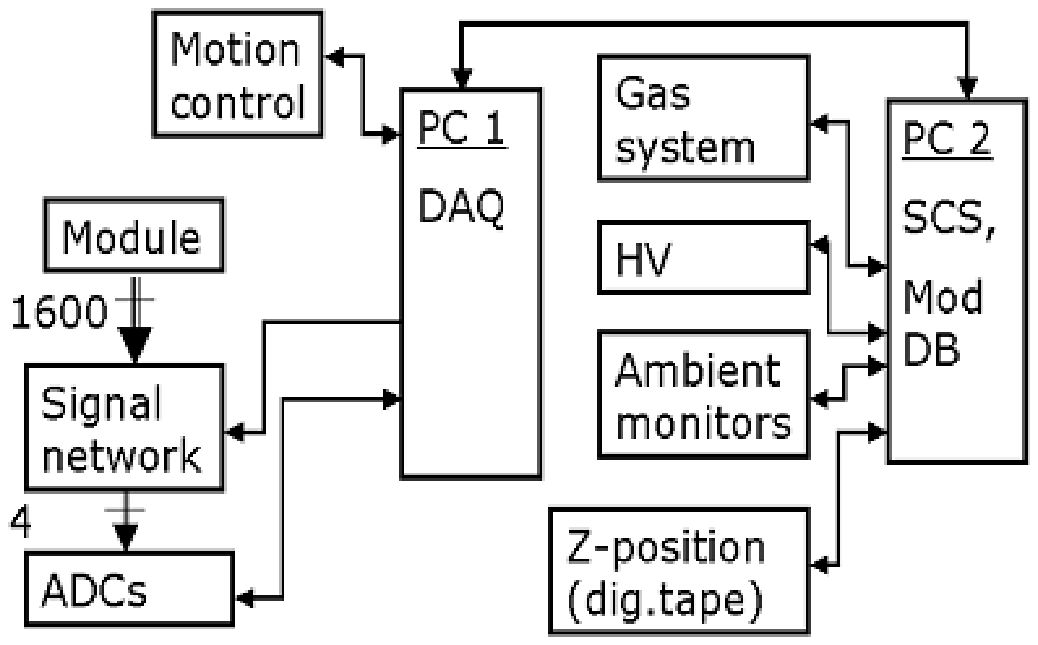}
    }
  }
 \caption{\it A simplified schematic of the X-ray scanner 
electronics and data acquisition scheme.
    \label{fig1} }
\end{figure}

\begin{figure}[hbtp]
  \centerline{
   \hbox{ 
     \hspace{0.1cm}
     \includegraphics[width=6.5cm, height=3.5cm]{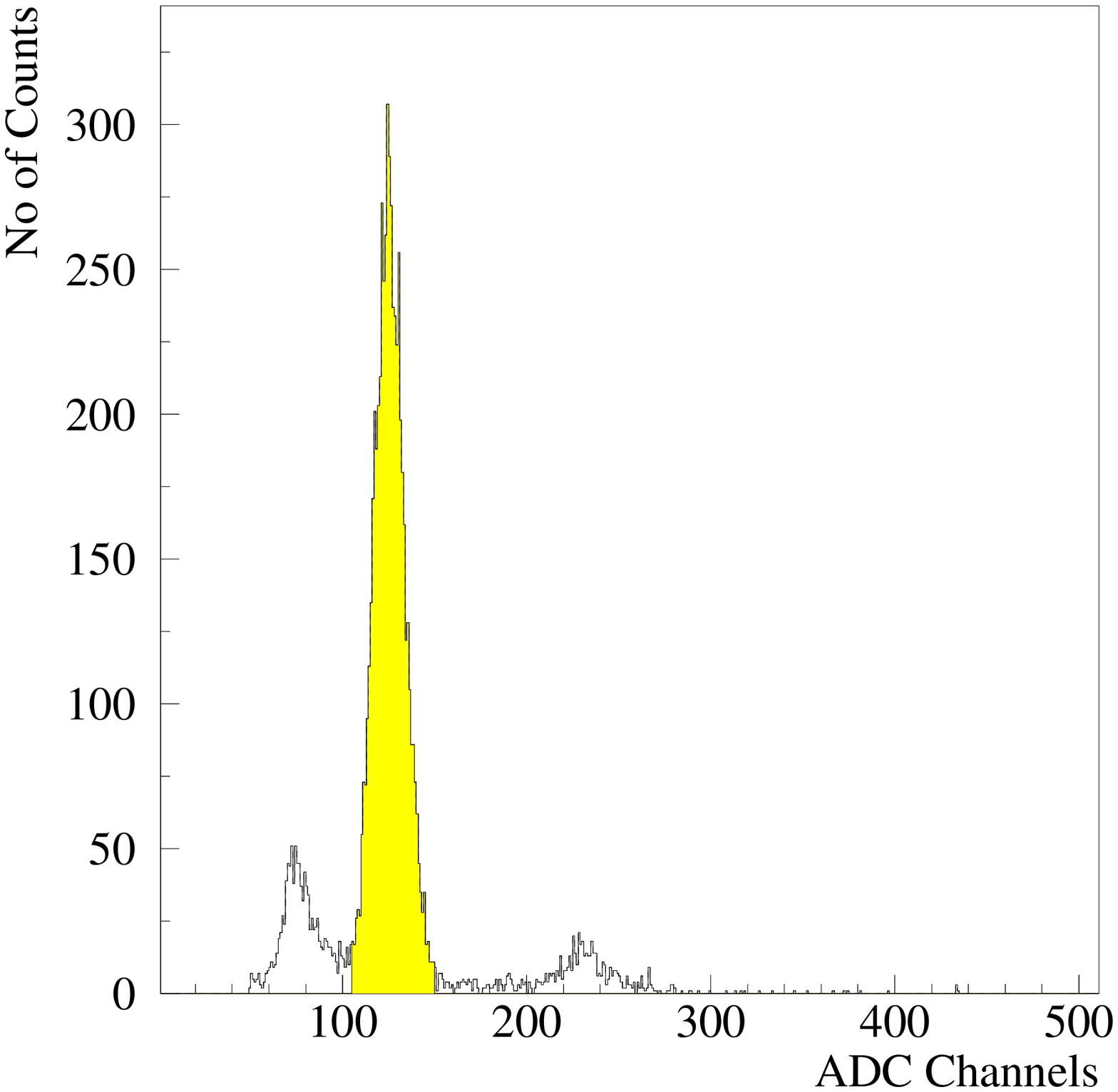}
     \hspace{0.1cm}
     \includegraphics[width=6.5cm, height=3.5cm]{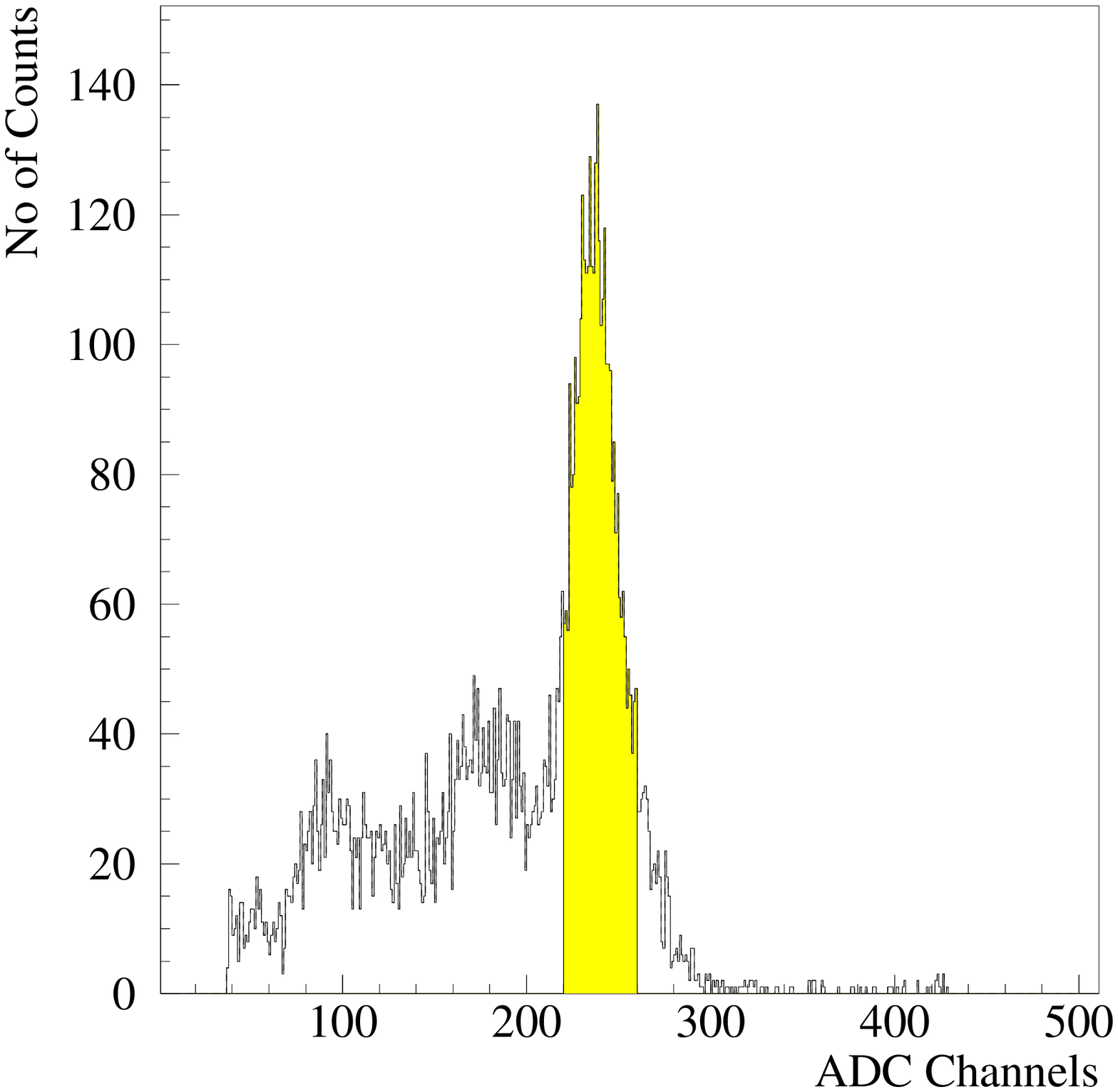}
    }
  }
 \caption{\it A typical spectrum from $^{55}$Fe in (a) and Bromine 
XRF in (b) from a straw.
    \label{fig2} }
\end{figure}

The basic chain of the electronics \cite{hu-ken} and the
data acquisition (DAQ) scheme are shown 
in figure \ref{fig1}. The signals from the sense wires are amplified and 
shaped and then passed on to ADCs. For each wire, spectra are collected from
50 positions along the $z$ direction.  For each spectrum, a monitor spectrum
was collected from a fixed point on a straw (not being scanned) exposed to 
an $^{55}$Fe source. The ionization gas was a mixture of 
argon and CO$_2$ (70:30) with a flow rate of about 1 straw volume/hr. 
The nominal operating high voltage (HV) for 
surveying was set to 1235 volts. Figure \ref{fig2} shows a typical spectrum 
from the $^{55}$Fe source and bromine XRF, respectively.

\begin{figure}[hbtp]
  \centerline{
   \hbox{ 
     \hspace{0.1cm}
     \includegraphics[width=6.5cm, height=4cm]{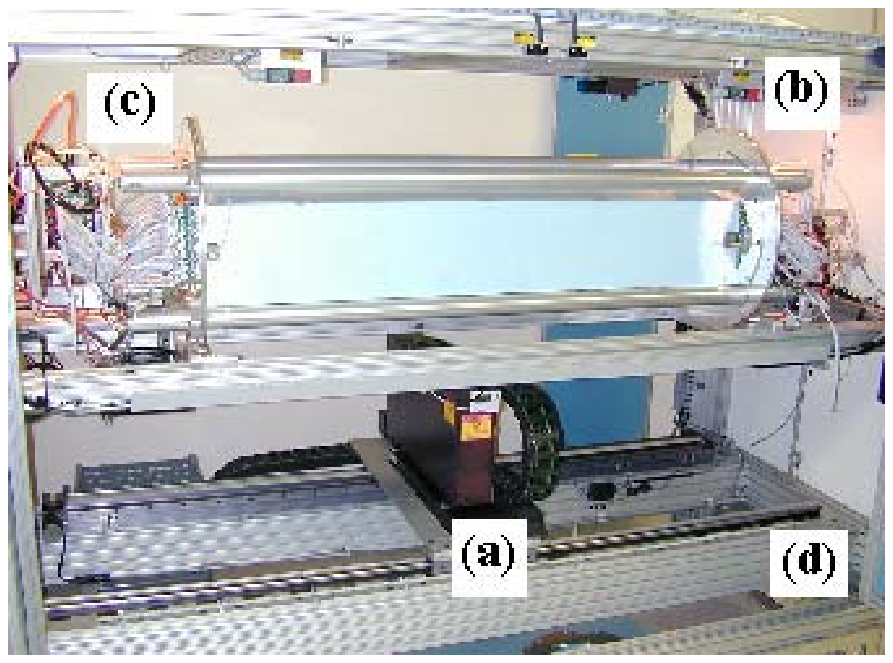}
     \hspace{0.1cm}
     \includegraphics[width=6.5cm, height=4cm]{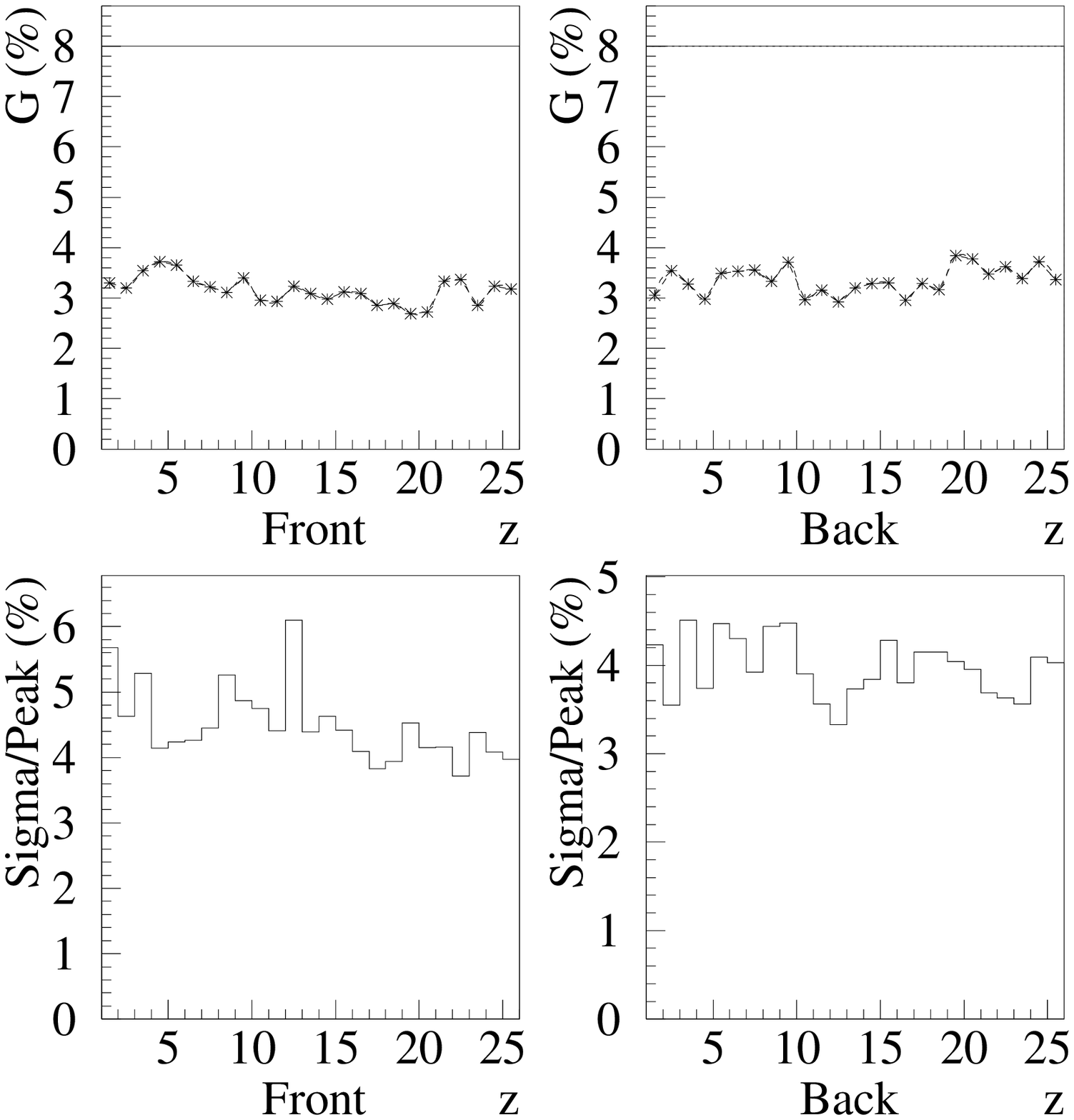}
    }
  }
 \caption{\it
  A picture of one X-ray scanner is shown with X-ray source (a) on the 
stepping motor rails (d) and front-end electronics (b and c). 
Also, a sample normalized gain variation $G$ along the $z$ direction from 
one straw is shown on the right. The solid line indicates $G$ of 8{\%}. 
Wires over 8{\%} of $G$ are flagged for possible restringing or removal.     
    \label{fig3} }
\end{figure}

The raw spectra were recorded with straw number, run number, $z$ position, and 
slow control data. Later, the peak in each spectrum was fitted with a Gaussian 
function that determines the mean ($g_p$), standard deviation ($\sigma_p$), and
goodness of fit. The ratio of the straw mean to the monitor mean of $^{55}$Fe 
is multiplied by 500 to give the normalized gain, $g_n$. The gain variation 
$G$ is defined as $G=(g_{n, max} - g_{n, min})/g_{n, min}$ and used to 
determine the wire offset from the center as shown in figure \ref{fig3}.  

\section{Summary}
X-ray scanners for the purpose of quality control of the ATLAS TRT Barrel 
detector modules were developed at Hampton University and are operating 
at HU and CERN. The scanners map the gain using 12 keV photons at 50 points 
along each straw. Results were used to decide whether wires were removed or 
re-strung, and to evaluate overall module quality.

\end{document}